\documentclass[aps,prb,twocolumn,superscriptaddress]{revtex4}

\usepackage{amsmath,amssymb}
\usepackage{graphicx}
\usepackage{bm}
\usepackage{epstopdf}
\usepackage{color}

\newcommand{\TcK}{$T_c^{\mathcal{K}}$}
\newcommand{\Tcac}{$T_c^\text{ac}$}

\begin{document}

\title{Anomalous superconducting state in LiFeAs implied by the $^{75}$As 
Knight shift measurement}

\author{S.-H. Baek}
\email[]{sbaek.fu@gmail.com}
\affiliation{IFW-Dresden, Institute for Solid State Research,
PF 270116, 01171 Dresden, Germany}
\author{L. Harnagea}
\affiliation{IFW-Dresden, Institute for Solid State Research,
PF 270116, 01171 Dresden, Germany}
\author{S. Wurmehl}
\affiliation{IFW-Dresden, Institute for Solid State Research,
PF 270116, 01171 Dresden, Germany}
\author{B. B\"{u}chner}
\affiliation{IFW-Dresden, Institute for Solid State Research,
PF 270116, 01171 Dresden, Germany}
\affiliation{Institut f\"ur Festk\"orperphysik, Technische Universit\"at 
Dresden, 01062 Dresden, Germany} 
\author{H.-J. Grafe}
\affiliation{IFW-Dresden, Institute for Solid State Research,
PF 270116, 01171 Dresden, Germany}
\date{\today}

\begin{abstract}
$^{75}$As NMR investigation of a single crystal of superconducting LiFeAs is 
presented.  The Knight shift and the \textit{in situ} ac susceptibility 
measurements as a function of temperature and external field are 
indicative of two 
superconducting (SC) transition temperatures, each of which is associated with its own 
upper critical field. Strikingly, the Knight shift maintains its normal state 
value over a temperature range in the SC state before it drops abruptly 
being consistent with spin-singlet pairing. 
Together with our previous NMR study, the anomalous SC 
state featured by the constant Knight shift is attributed to the extremely 
sensitive SC properties of LiFeAs, probably stemming from its proximity to a 
critical instability.  
\end{abstract}


\maketitle

It is commonly argued that superconductivity in iron pnictides is driven by the 
antiferromagnetic (AFM) spin fluctuations which  
are associated with nesting between the hole and electron Fermi surface pockets, although 
the SC gap symmetry seems to vary among the materials from 
nodal to nodeless.\cite{lumsden10a,stewart11,chubukov12}
An exception in this general picture is LiFeAs that is superconducting as is, without 
any signature of nesting and static magnetism, yet with rather high $T_c\sim18$ 
K.\cite{morozov10,borisenko10,pitcher10}

While the absence of nesting and static magnetism in LiFeAs\cite{borisenko10} 
might support a non-magnetic origin for the SC pairing, such as phonons\cite{kordyuk11} or 
orbital fluctuations,\cite{kontani10} AFM spin fluctuations remain 
a strong candidate that is responsible  
for the SC pairing,\cite{jeglic10,platt11,taylor11,hajiri12} e.g., by 
recovering the nesting condition by the magnetic response shifting.\cite{qureshi12} 
If this is indeed the case, it would strengthen the 
belief that AFM spin fluctuations are fundamental to the superconductivity 
of iron-pnictides.  
On the other  
hand, spin-triplet pairing which is driven by ferromagnetic spin fluctuations 
originating from strong Hund coupling was also suggested,\cite{brydon11}
being followed by some experimental supports.\cite{pramanik11,haenke12,baek12,li13}
Such debates about the pairing mechanism in LiFeAs may imply that the nature 
of superconductivity in this material  
is different from other iron-pnictide families, and it was suggested 
that close proximity of the system to a strong magnetic instability may effect
the unusual sensitivity of the SC properties.\cite{baek12,brydon11}

In an effort to confirm the 
underlying instability and to uncover its nature, we carried out $^{75}$As nuclear 
magnetic resonance (NMR) in a single crystal of LiFeAs chosen from a different batch 
than those used in our previous NMR study,\cite{baek12} focusing on the low 
temperature range near and below $T_c\sim 18$ K. 
While the \textit{in situ} ac susceptibility and the NMR signal intensity 
confirm the bulk $T_c$, the Knight shift remains constant down to a temperature
at which it drops sharply. Although the constant Knight shift behavior in the 
SC state is not easily reproducible in other single crystals, our data suggest 
that an anomalous superconducting state where the Knight shift does not change could be 
stabilized. We discuss that LiFeAs is very close to a critical instability  
which affects the SC state particularly near the region of the 
normal/superconducting boundary.

The single crystal of LiFeAs was grown by a self-flux method as described in 
Ref. \onlinecite{morozov10}. Due to the sensitivity of the sample to air and moisture, 
the sample was carefully sealed  
into a quartz tube filled with Ar gas for NMR measurements.  
The sealed sample was 
mounted on a goniometer for an accurate alignment of the sample along 
the external field. 
$^{75}$As ($I=3/2$) nuclear magnetic resonance (NMR) experiments were 
performed in the range of temperature  
3.6 --- 25 K and external field 0 --- 16 T.  We also carried out $^{75}$As nuclear 
quadrupole resonance (NQR) to determine the quadrupole frequency $\nu_Q$. The 
NQR spectrum shows a width of 75 kHz at 20 K, which is much narrower than 
$\sim 170$ kHz in a powder sample\cite{li10} and thus indicates a sign of good
chemical homogeneity of the sample. 

The Knight shift $\mathcal{K}$, i.e., the local static spin susceptibility, 
was measured from
$^{75}$As NMR central line at 
various external fields applied parallel and perpendicular to the crystallographic $c$ axis. 
The large quadrupole frequency $\nu_Q=21.08$ MHz of the  
$^{75}$As ($I=3/2$), which is almost $T$-independent in the low temperature range 
investigated, shifts the central transition of  
the $^{75}$As by the second order quadrupole effect given by 
$\Delta\nu=3\nu_Q^2/16\omega_n(1-\cos^2\theta)(1-9\cos^2\theta)$ for $I=3/2$ 
where $\omega_n$ is the unshifted Larmor frequency and $\theta$ is the angle 
between the external field $H$ and the $c$ axis.
The Knight shift shown in Fig. 1 was obtained by 
subtracting $\Delta\nu$ from the total shift of the central line. The 
SC transition temperature $T_c$ was identified  
from a sudden drop of $\mathcal{K}$ for a given external field $H$, which indicates  
spin singlet Cooper pairing. Whereas this behavior seems consistent with 
previous other NMR studies in this compound,\cite{li10,jeglic10} we find that 
$T_c(H)$, particularly  
for $H \parallel c$, are much lower than values reported in 
literature.\cite{lee10a,heyer11,kurita11,cho11,khim11}  At 8.5 T for $H\parallel c$, for 
example, $\mathcal{K}$ does not drop down even at  
3.6 K, indicating that $T_c\le3.5$ K is significantly lower than an 
expected value ($>10$ K).\cite{lee10a,heyer11,kurita11,cho11,khim11}

\begin{figure}
\centering
\includegraphics[width=\linewidth]{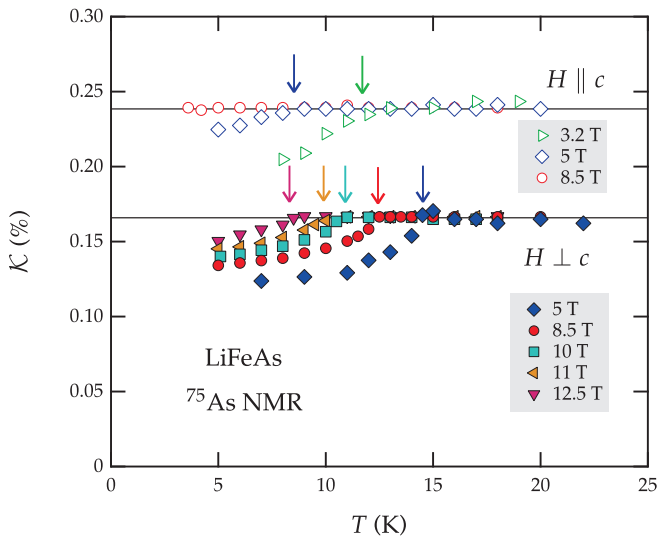}
\caption{\label{fig:knight} Knight shift ($\mathcal{K}$) of $^{75}$As 
as a function of temperature and field for two field orientations. A second order 
quadrupole correction was made for $H \perp c$.  Superconducting transition 
temperature for each field was determined from the sharp drop of $\mathcal{K}$ 
as denoted by arrows. 
}
\end{figure}

In order to confirm the transition temperature, we measured 
the \textit{in situ} ac susceptibility $\chi_\text{ac}$ using the NMR radio 
frequency (rf) circuit. In the SC 
state, the Meissner effect induces the change of impedance and 
thus the tuning frequency of the rf circuit. Therefore, the onset of 
superconductivity could be  
detected by monitoring $\chi_\text{ac}$ as a function of 
temperature. Fig. 2 shows $\chi_\text{ac}(T)$ measured at various external 
fields $H$.  
Here we define $T_c$ as  
a temperature where $\chi_\text{ac}$ reaches 10\% of the full drop to the low temperature 
plateau at each field, which are denoted by down  
arrows. Clearly $T_{c}$ detected by $\chi_\text{ac}$ is much 
higher than that obtained 
by the Knight shift measurements for each field (up arrows).  
Note that, at 8.5 T parallel to the $c$ axis, a clear  
onset was observed at $11$ K by $\chi_\text{ac}$ which is compatible with 
values reported thus far,\cite{lee10a,heyer11,kurita11,cho11,khim11} in stark 
contrast to the absence of the Knight shift anomaly down to 3.6 K.
It may be worthwhile to note that $\chi_\text{ac}$ displays a small but noticeable anomalous 
change in its slope at $T_c$ obtained by $\mathcal{K}$.  

As the two experimental methods seem to distinguish different onset temperatures of 
the SC transition, here we define the two onset temperatures obtained by 
$\chi_\text{ac}$ and $\mathcal{K}$ by \Tcac\ and \TcK, respectively.   
While a sharp drop of $\mathcal{K}$ is usually a good indication of spin 
singlet superconductivity, $\chi_\text{ac}$  
alone is not sufficient in general to verify a bulk $T_c$, because other 
non-superconducting effects might alter the  
temperature dependence of $\chi_\text{ac}$. 
To check the validity of \Tcac, we carefully examined the 
temperature evolution of the $^{75}$As spectra. In the  
SC state, the signal intensity should decrease due to 
supercurrents which reduce the sample volume that can be penetrated by the rf 
field, and therefore it could be another good probe for detecting the onset of 
bulk superconductivity.
Fig. 3 shows the $^{75}$As NMR spectrum as a function of temperature measured at 
8.5 T, where the Boltzmann correction was  
made by multiplying $T$ for each spectrum.  For $H\perp c$, 
the signal intensity starts to decrease at $\sim 15$ K, which agrees with 
\Tcac\ determined from   
$\chi_\text{ac}$, as shown in Fig. 3(c). The agreement of the 
signal intensity with  
$\chi_\text{ac}$ in their temperature dependences was also confirmed for 
$H \parallel c$ [see Fig. 3(b) and (c)]. Note that  
the anisotropy of $\chi_\text{ac}$ below \Tcac\ 
remarkably coincides with that of the signal intensity.
For direct comparison, 
$\mathcal{K}$ is shown in the upper panel of Fig. 3(c). 

\begin{figure}
\centering
\includegraphics[width=\linewidth]{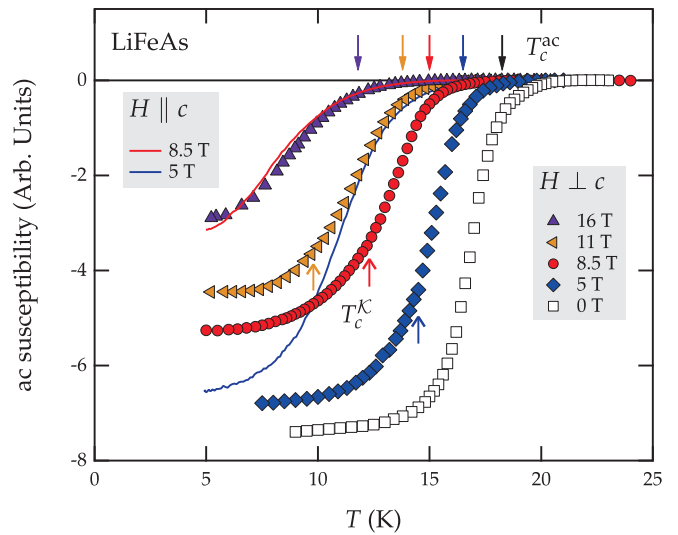}
\caption{\label{fig:chiac} \textit{In situ} ac susceptibility $\chi_\text{ac}$ measured 
in the NMR tank circuit as a function of temperature and external field. 
The transition temperature \Tcac\ (down arrows) is considerably higher than 
\TcK\ (up arrows) determined by the  
Knight shift measurements. Data for $H \parallel c$ are shown as solid lines 
(no arrows for clarity).  } 
\end{figure}

We emphasize that $T_c^\text{ac}(H)$ indeed confirms bulk 
superconductivity which has been unanimously proven in our single crystals by 
numerous other measurements including dc magnetic  
susceptibility,\cite{morozov10} specific heat,\cite{stockert11} resistivity,\cite{heyer11} 
angle-resolved photoemission spectroscopy 
(ARPES),\cite{borisenko10,kordyuk11,borisenko12} neutron,\cite{qureshi12} and 
scanning tunneling spectroscopy (STS),\cite{haenke12} as well  
as by theoretical supports.\cite{lankau10,knolle12}
In particular, note that the  
specific heat measured in our single crystals manifests the   
bulk $T_c\sim 9$ K in a field of 9 T applied along  
the $c$ axis\cite{stockert11} that is comparable to \Tcac\ at 8.5 T, while 
the Knight shift remains constant down to 3.6 K at 8.5 T (see Fig. 3). 
Furthermore, $T_c^\text{ac}(H)$ and the related $H_{c2}$  
are in satisfactory agreement with the results measured in other samples 
by different groups.\cite{kurita11,cho11,khim11}
Therefore, we conclude that \Tcac\ is equivalent to the onset temperature of 
the \textit{bulk} Meissner effect which modifies the signal intensity  
and $\chi_\text{ac}$ simultaneously. 

Further analysis of the Knight shift, the signal intensity, and the 
ac susceptibility obtained at various external fields reveals quite different 
field dependence  of \TcK\ and \Tcac. (For 
raw $^{75}$As spectra at external fields other than 8.5 T, see supplemental 
material.\cite{supple}) 
The resulting $H$-$T$ phase 
diagram is presented in Fig. 4. We find that the $H$-dependence of \Tcac\ is 
in qualitative agreement with 
other studies.\cite{heyer11, khim11,kurita11} For example, the 
data from Khim et al.\cite{khim11} are compatible with 
\Tcac\ data in the $H$-dependence as well as the anisotropy.

\begin{figure}
\centering
\includegraphics[width=\linewidth]{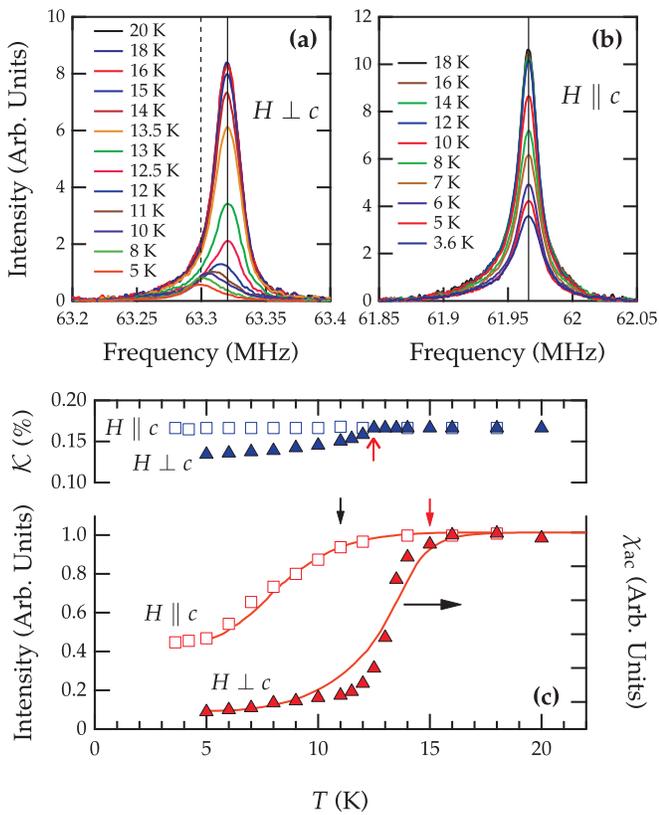}
\caption{\label{fig:sp_comp}Temperature dependence of $^{75}$As NMR central line 
at $H=8.5$ T for (a) $H\perp c$ and (b) $H \parallel c$. (c) Signal 
intensity and Knight shift versus temperature at 8.5 T. Temperature 
dependence of signal intensity for both $H \perp c$ and $H\parallel c$ agrees well
with that of $\chi_\text{ac}$, indicating that \Tcac\  
represents the onset of screening due to superconductivity. The Knight shift, 
however, reveals \TcK\ which is significantly lower than \Tcac. 
$\mathcal{K}_\parallel$ was offset vertically for comparison.   
}
\end{figure}

In contrast, the $H$-dependence of \TcK\ is very different from that of 
\Tcac, other than its much lower values.  
For $H \perp c$, while \Tcac\ exhibits almost a linear $H$-dependence up to 16 T,
\TcK\ does not decrease linearly with increasing $H$.  
Consequently, the difference $T_c^\text{ac}-T_c^\mathcal{K}$ becomes 
larger at higher fields. This trend is more pronounced for $H \parallel c$. 
Note that the estimated $H$-dependence of \TcK\ for 
$H\parallel c$ (dashed line in Fig. 4) agrees with the absence of \TcK\ at  
8.5 T down to 3.6 K. 

\begin{figure}
\centering
\includegraphics[width=\linewidth]{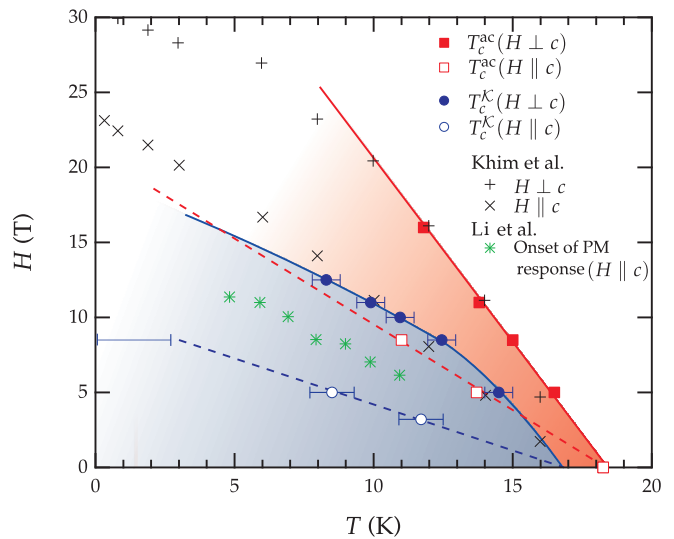}
\caption{\label{fig:phasedia} $H$-$T$ phase diagram in LiFeAs. Two 
onset temperatures \Tcac\ and \TcK\ were 
obtained by the ac susceptibility and 
the Knight shift, respectively. Data from Khim et al.\cite{khim11} and Li et 
al.\cite{li13} are shown for comparison. The onset  
temperatures of the paramagnetic irreversibility from Li et al. fall between  
\Tcac\ and \TcK\ line, while $T_c (H\parallel c)$ 
in Ref. \onlinecite{li13} determined from  
resistivity (not shown for clarity) almost coincides with $T_c^\text{ac} (H)$. 
Note that shades of blue and red are applicable only to the 
case of $H \perp c$ and thus some care is needed to compare the data for 
$H\parallel c$.}  
\end{figure}

Our experimental results naturally raise important questions.  Does 
$\mathcal{K}$, i.e., the intrinsic spin susceptibility, remain unchanged across 
\Tcac\ but drop   
below \TcK? Do the two seemingly distinguishable SC states above and below \TcK occur 
in a single phase?  
It should be emphasized that only one 
phase must be present in the normal state above \Tcac, because NMR and NQR 
spectra exhibit very sharp single lines, and their signal intensities are well
conserved at all temperatures investigated. 
Although inhomogeneous superconductivity is extremely unlikely due to 
bulk superconductivity in our single crystals, here we discuss the 
possibility that the two SC transitions result from \textit{phase segregation} 
in bulk form below \Tcac, i.e., a partial volume fraction of  
the sample (region I) becomes superconducting at \Tcac\ first, and the rest of 
the sample (region II) remains normal down to \TcK\ but undergoes the SC
transition at \TcK. 

If phase segregation takes place at \Tcac, the otherwise 
single spectrum would be segregated into two parts arising from SC region I 
and normal region II, respectively.  
In this case, the unchanged Knight shift of the 
``total'' spectrum between \TcK and \Tcac\ 
could be realized \textit{only} either (i) if the SC transition in region I is 
extremely sharp so that  
the decreasing Knight shift is not detected, or (ii) if triplet 
superconductivity occurs in region I so that $\mathcal{K}$ of region I is 
still the same as that of the normal region II.   
The consequence of case (i) should be an almost discontinuous 
change of the signal  
intensity just below \Tcac. On the contrary, we find that the $T$-dependence of the   
signal intensity shows a gradual change over a temperature range [see Fig. 
3(c)], ruling out this scenario.  
Similarly, case (ii) is also ruled out as following.  
Since singlet superconductivity occurs at \TcK, we should have two 
different SC pairing states below \TcK. Since $\mathcal{K}$ from region II 
decreases while $\mathcal{K}$ from region I remains constant, two NMR lines or 
noticeable broadening below \TcK\ should be observed.  As shown in Fig. 3(a), however, 
the well-defined single line at all temperatures is inconsistent  
with this scenario.
Also by a close inspection of the $T$-dependence 
of $^{75}$As spectrum at other external fields,\cite{supple} the phase segregation 
scenario turns out highly improbable. Hence, we reach the remarkable conclusion 
that $\mathcal{K}$ is indeed a constant in the SC state between 
\Tcac\ and \TcK, suggesting that the anomalous SC state may change at \TcK\ to 
a somewhat ``normal'' SC state with singlet-pairing symmetry.

\textit{A priori}, the unchanged Knight shift through 
\Tcac\ contrasts with spin-singlet pairing, because it implies that 
the spin degree of freedom of electrons does not vanish in the SC state. 
Surprisingly, another signature of the possible unusual SC state in LiFeAs was 
also verified independently in a different single crystal by recent magnetometry  
measurements\cite{li13} which report a
paramagnetic (PM) response within the SC state at high fields.  The onset temperature of 
PM irreversibility $T_\text{irr}$ as a 
function of $H \parallel c$ is located between the
\Tcac\ and \TcK\ lines (see Fig. 4), whereas $T_c$ determined from resistivity is well
consistent with $T_c^\text{ac}(H \parallel c)$. The anomalous PM response in the SC state, 
which is ascribed to   
the triplet component induced by high fields,\cite{li13} is indeed in excellent 
agreement with the nonvanishing spin susceptibility in the SC state revealed 
by the constant Knight shift. Note that,  
since $T_\text{irr}$ is a  
crossover temperature rather than a measure of the actual transition, 
$T_\text{irr}>T_{c}^\mathcal{K}$ is very reasonable. Therefore, combining our 
NMR results and Ref.  
\onlinecite{li13}, we interpret that both the 
constant Knight shift and the PM response observed in the similar region of the 
phase diagram are signs of spin-triplet pairing that could perhaps be stabilized under 
certain conditions, which may be a realization of the theoretical prediction that a 
spin-triplet could occur in iron-pnictides depending on 
various parameters such as Hund  
coupling and onsite Coulomb repulsion.\cite{daghofer08,nicholson11,brydon11}

Interestingly, the PM irreversibility at high  
fields was not reproduced in other samples, being attributed to 
the extreme sensitivity of samples\cite{li13} which was already proposed in our 
previous NMR study.\cite{baek12}
Such a difficult reproducibility of the constant Knight shift behavior 
as well as of the PM response in Ref. \onlinecite{li13}, together with the extreme 
sensitivity of the SC properties demonstrated in our previous NMR 
study,\cite{baek12} suggests that the   
triplet-like anomalous SC state is unstable 
in nature, being susceptible to even a tiny  
off-stoichiometry. This is also consistent with a large 
variation of $T_c$ from 15.5 K to 18 K which has been found in a recent 
transport study,\cite{rullier12} although all the measured samples show 
the lowest residual-resistivity   
ratios and thus appear to be of high quality.

Here we argue that the peculiar sensitivity of LiFeAs could be a 
natural consequence of the close proximity to a  
critical instability, near which the unusual SC state could emerge. 
Hence, as long as the off-stoichiometry is small (i.e., the sample quality is pure 
enough), the effect of the instability would persist especially at 
high temperatures/fields  
causing $T_c (H=0)$ and $H_{c2} (T=0)$ very much sample-dependent, whether or 
not the unusual SC state is actually stabilized.  
Note that this picture indeed accounts well for
the non-trivial large variation of $T_c$ 
and $H_{c2}$ reported so far in LiFeAs\cite{heyer11,kurita11,cho11,khim11,rullier12}
and the persisting 2D superconducting fluctuations up to 1.4$T_c$.\cite{rullier12}
Furthermore, given the possible realization of the anomalous SC state 
which differs from the usual spin-singlet state, 
contradicting experimental results 
regarding the pairing symmetries in LiFeAs\cite{haenke12,hashimoto12,jang12} may 
be reconciled with each other, in terms of the closeness to a critical 
ferromagnetic instability.  


This work has been
supported by the Deutsche Forschungsgemeinschaft through SPP1458 (Grant No. GR3330/2 and
BE1749/13) and through Emmy Noether Programme WU595/3-1.

\bibliography{mybib}

\newpage

\begin{figure}
\centering
\includegraphics[width=\textwidth]{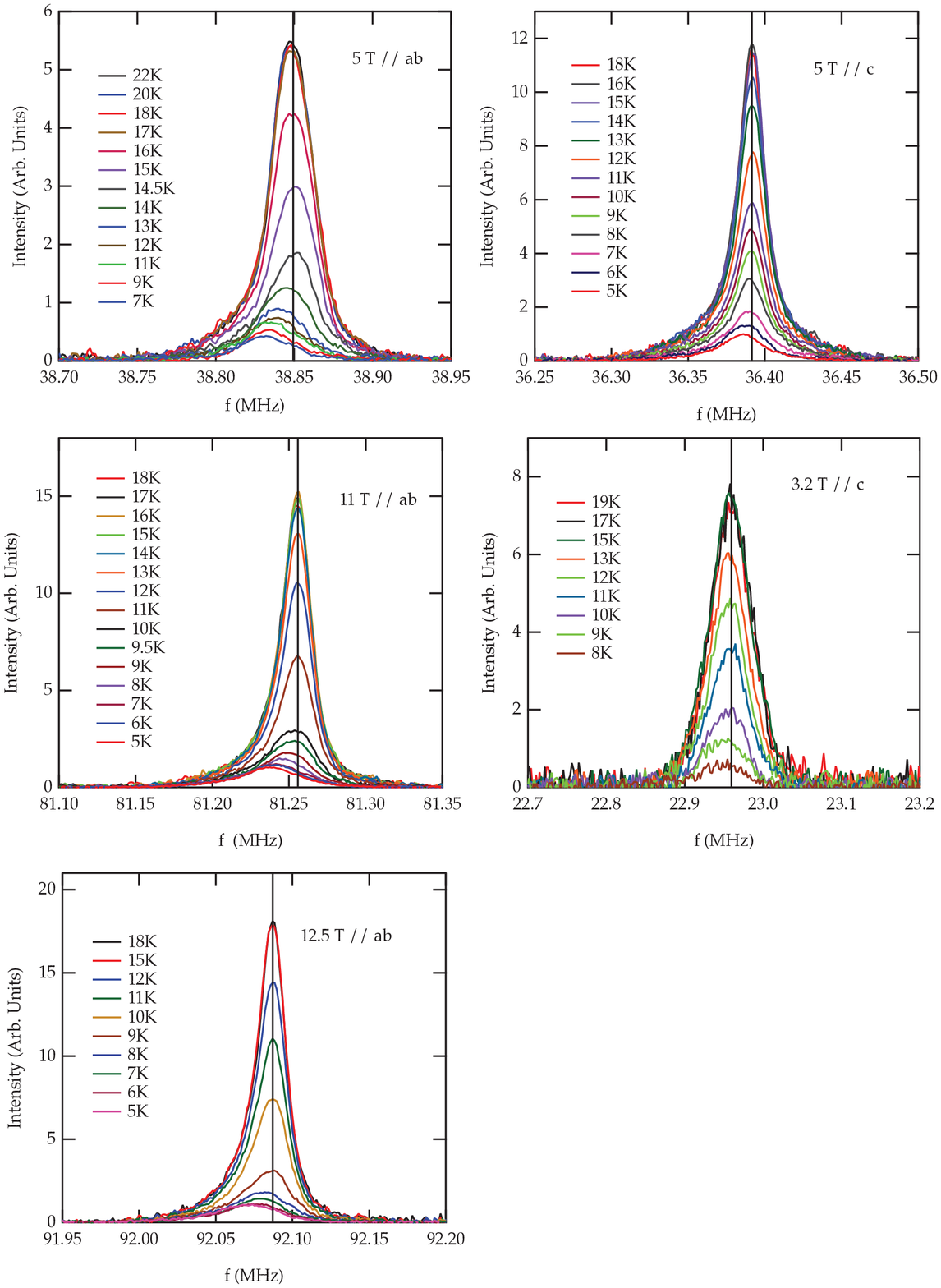}
\caption{Supplemental figure}
\end{figure}

\end{document}